\PassOptionsToPackage{unicode}{hyperref}
\PassOptionsToPackage{hyphens}{url}
\PassOptionsToPackage{dvipsnames,svgnames,x11names}{xcolor}

\documentclass[
]{article}
\usepackage[a4paper, total={6.5in, 10in}]{geometry}

\usepackage{amsmath,amssymb}
\usepackage{lmodern}
\usepackage{iftex}
\ifPDFTeX
  \usepackage[T1]{fontenc}
  \usepackage[utf8]{inputenc}
  \usepackage{textcomp} 
\else 
  \usepackage{unicode-math}
  \defaultfontfeatures{Scale=MatchLowercase}
  \defaultfontfeatures[\rmfamily]{Ligatures=TeX,Scale=1}
\fi
\IfFileExists{upquote.sty}{\usepackage{upquote}}{}
\IfFileExists{microtype.sty}{
  \usepackage[]{microtype}
  \UseMicrotypeSet[protrusion]{basicmath} 
}{}
\makeatletter
\@ifundefined{KOMAClassName}{
  \IfFileExists{parskip.sty}{%
    \usepackage{parskip}
  }{
    \setlength{\parindent}{0pt}
    \setlength{\parskip}{6pt plus 2pt minus 1pt}}
}{
  \KOMAoptions{parskip=half}}
\makeatother
\usepackage{xcolor}
\usepackage{graphicx}
\makeatletter
\def\maxwidth{\ifdim\Gin@nat@width>\linewidth\linewidth\else\Gin@nat@width\fi}
\def\maxheight{\ifdim\Gin@nat@height>\textheight\textheight\else\Gin@nat@height\fi}
\makeatother
\setkeys{Gin}{width=\maxwidth,height=\maxheight,keepaspectratio}
\makeatletter
\def\fps@figure{htbp}
\makeatother
\NewDocumentCommand\citeproctext{}{}
\NewDocumentCommand\citeproc{mm}{%
  \begingroup\def\citeproctext{#2}\cite{#1}\endgroup}
\makeatletter
 \let\@cite@ofmt\@firstofone
 \def\@biblabel#1{}
 \def\@cite#1#2{{#1\if@tempswa , #2\fi}}
\makeatother
\newlength{\cslhangindent}
\newlength{\csllabelwidth}
\newenvironment{CSLReferences}[2] 
 {\begin{list}{}{%
  \setlength{\itemindent}{0pt}
  \setlength{\leftmargin}{0pt}
  \setlength{\parsep}{0pt}
  \ifodd #1
   \setlength{\leftmargin}{\cslhangindent}
   \setlength{\itemindent}{-1\cslhangindent}
  \fi
  \setlength{\itemsep}{#2\baselineskip}}}
 {\end{list}}
\usepackage{calc}

\ifLuaTeX
\usepackage[bidi=basic]{babel}
\else
\usepackage[bidi=default]{babel}
\fi
\babelprovide[main,import]{american}

\def\languageshorthands#1{}
\ifLuaTeX
  \usepackage{selnolig}  
\fi
\IfFileExists{bookmark.sty}{\usepackage{bookmark}}{\usepackage{hyperref}}
\IfFileExists{xurl.sty}{\usepackage{xurl}}{} 
\hypersetup{
  pdftitle={DAXA: Traversing the X-ray desert by Democratising Archival
X-ray Astronomy},
  pdfauthor={David J. Turner, Jessica E. Pilling, Megan Donahue, Paul A.
Giles, Kathy Romer, Agrim Gupta, Toby Wallage, Ray Wang},
  pdflang={en-US},
  colorlinks=true,
  linkcolor={Blue},
  filecolor={Maroon},
  citecolor={Blue},
  urlcolor={Blue},
  pdfcreator={LaTeX via pandoc}}

\title{DAXA: Traversing the X-ray desert by Democratising Archival X-ray
Astronomy}

\definecolor{c53baa1}{RGB}{83,186,161}
\definecolor{c202826}{RGB}{32,40,38}

\renewcommand*{\thefootnote}{\fnsymbol{footnote}}

\usepackage[affil-it]{authblk}
\usepackage{orcidlink}
\author[1,2%
  ]{David J. Turner\footnote{turne540@msu.edu}%
    \,\orcidlink{0000-0001-9658-1396}\,%
    }
\author[2%
  ]{Jessica E. Pilling%
    \,\orcidlink{0000-0002-3211-928X}\,%
    }
\author[1%
  ]{Megan Donahue%
    \,\orcidlink{0000-0002-2808-0853}\,%
    }
\author[2%
  ]{Paul A. Giles%
    \,\orcidlink{0000-0003-4937-8453}\,%
    }
\author[2%
  ]{Kathy Romer%
    \,\orcidlink{0000-0002-9328-879X}\,%
    }
\author[1%
  ]{Agrim Gupta%
    }
\author[2%
  ]{Toby Wallage%
    }
\author[1%
  ]{Ray Wang%
    \,\orcidlink{0000-0003-2102-8646}\,%
    }

\affil[1]{Department of Physics and Astronomy, Michigan State
University, Lansing, Michigan, USA%
  }
\affil[2]{Department of Physics and Astronomy, University of Sussex,
Brighton, UK%
  }
\date{}

\begin{document}
\maketitle

\renewcommand*{\thefootnote}{\arabic{footnote}}
\setcounter{footnote}{0}

\section{Summary}\label{summary}

We introduce a new, open-source, Python module for the acquisition and
processing of archival data from many X-ray telescopes, Democratising
Archival X-ray Astronomy (hereafter referred to as \textsc{Daxa}\footnote{\href{https://github.com/DavidT3/DAXA}{{\sc Daxa} GitHub - https://github.com/DavidT3/DAXA}}). The
aim of \textsc{Daxa} is to provide a unified, easy-to-use, and fully documented\footnote{\href{https://daxa.readthedocs.io/}{{\sc Daxa} Documentation - https://daxa.readthedocs.io/}} Python
interface to the disparate X-ray telescope data archives and their
processing tools. We provide this interface for the majority of X-ray
telescopes launched within the last 30 years. This module enables much
greater access to X-ray data for non-specialists, while preserving
low-level control of processing for X-ray experts. It is useful for
identifying relevant observations of a single object of interest, but it
excels at creating and managing multi-mission datasets for serendipitous
or targeted studies of large samples of X-ray emitting objects. Once
relevant observations are identified, the raw data can be downloaded
(and optionally processed) through \textsc{Daxa}, or pre-processed event
lists, images, and exposure maps can be downloaded if they are
available. With a decade-long `X-ray desert' potentially on the horizon,
archival data will take on even greater importance, and enhanced access
to those archives will be vital to the continuation of X-ray astronomy.

\section{Statement of need}\label{statement-of-need}

X-ray observations provide a powerful view of some of the most extreme
processes in the Universe, and have had a profound impact on our
understanding of many types of astrophysical objects. Every sub-field of
astronomy, astrophysics, and cosmology has benefited significantly from
X-ray coverage over the last three decades; the observation of X-ray
cavities in galaxy clusters caused by central AGN helped to shed light
on the cooling-flow problem (\citeproc{ref-cavities}{McNamara et al.,
2001}); further X-ray observations allowed for the measurement of
spatially-resolved entropy in hundreds of clusters, dramatically
increasing understanding of cooling and heating processes in their
cores; quasi-periodic eruptions (QPE) from active galactic nuclei (AGN)
were discovered (\citeproc{ref-qpedisco}{Miniutti et al., 2019}); the
high-energy view of young stars gave insights into their magnetic fields
and stellar winds (\citeproc{ref-coup}{Getman et al., 2005};
\citeproc{ref-xest}{Güdel et al., 2007}); calibrating mis-centering for
galaxy cluster weak-lensing studies helped constrain cosmological
parameters (\citeproc{ref-miscen}{Zhang et al., 2019}); and X-rays even
helped probe the irradiation of exoplanets
(\citeproc{ref-xrayirrexo}{Poppenhaeger et al., 2021}). Indeed, X-ray
telescopes have created many entirely new fields of study; they provided
the first evidence of X-ray sources outside the solar system
(\citeproc{ref-theOG}{Giacconi et al., 1962}); discovered the first
widely accepted black hole, and launched the study of supernova remnants
(\citeproc{ref-cygx1andfriends}{Bowyer et al., 1965}); and found
ionized, volume-filling, gas within the Coma galaxy cluster (the
intra-cluster medium) (\citeproc{ref-clusterdisco}{Kellogg et al.,
1971}), with the implication that clusters were more than collections of
galaxies. These non-exhaustive lists make evident the importance of
X-ray observations to the astronomy, astrophysics, and cosmology
communities.

The current workhorse X-ray observatories, \emph{XMM}
(\citeproc{ref-xmm}{Jansen et al., 2001}) and \emph{Chandra} are ageing,
with \emph{Chandra} in particular experiencing a decline in low-energy
sensitivity that might limit science cases (other telescopes are online
but are more specialised); these missions cannot last forever. If we are
to enter an X-ray desert, where the astrophysics community has only
limited access to new X-ray observations from specialised missions like
\emph{Swift} (\citeproc{ref-swift}{Gehrels et al., 2004}), \emph{NuSTAR}
(\citeproc{ref-nustar}{Harrison et al., 2013}), and \emph{XRISM}
(\citeproc{ref-xrism}{XRISM Science Team, 2020}), then archival data
(and serendipitous studies) take on an even greater value than they
already hold. \textsc{Daxa} is part of an ecosystem of open-source
software designed around the concept of enabling serendipitous studies
of X-ray emitting objects, and can download and prepare X-ray
observations for use with tools like `X-ray: Generate and Analyse'
(\textsc{Xga}; Turner et al. (\citeproc{ref-xga}{2022})). X-ray
observations are uniquely well suited for the kind of archival study
facilitated by \textsc{Daxa} and \textsc{Xga}, as they generally record
the time, position, and energy of each individual photon impacting the
detector (true for all missions currently implemented in \textsc{Daxa});
this means that we can create images, lightcurves, and spectra for any
object detected within the field-of-view, even if it was not the target.
With this software, we enable the maximum use of existing X-ray
archives, to traverse the X-ray desert and ensure that we are fully
prepared for future X-ray telescopes such as \emph{Athena}
(\citeproc{ref-athena}{Nandra et al., 2013}) and \emph{Lynx}
(\citeproc{ref-lynx}{Gaskin et al., 2019}). Having easy access to the
whole X-ray observation history of an object can provide valuable
astrophysical context at little extra cost.

\begin{figure}
\centering
\includegraphics{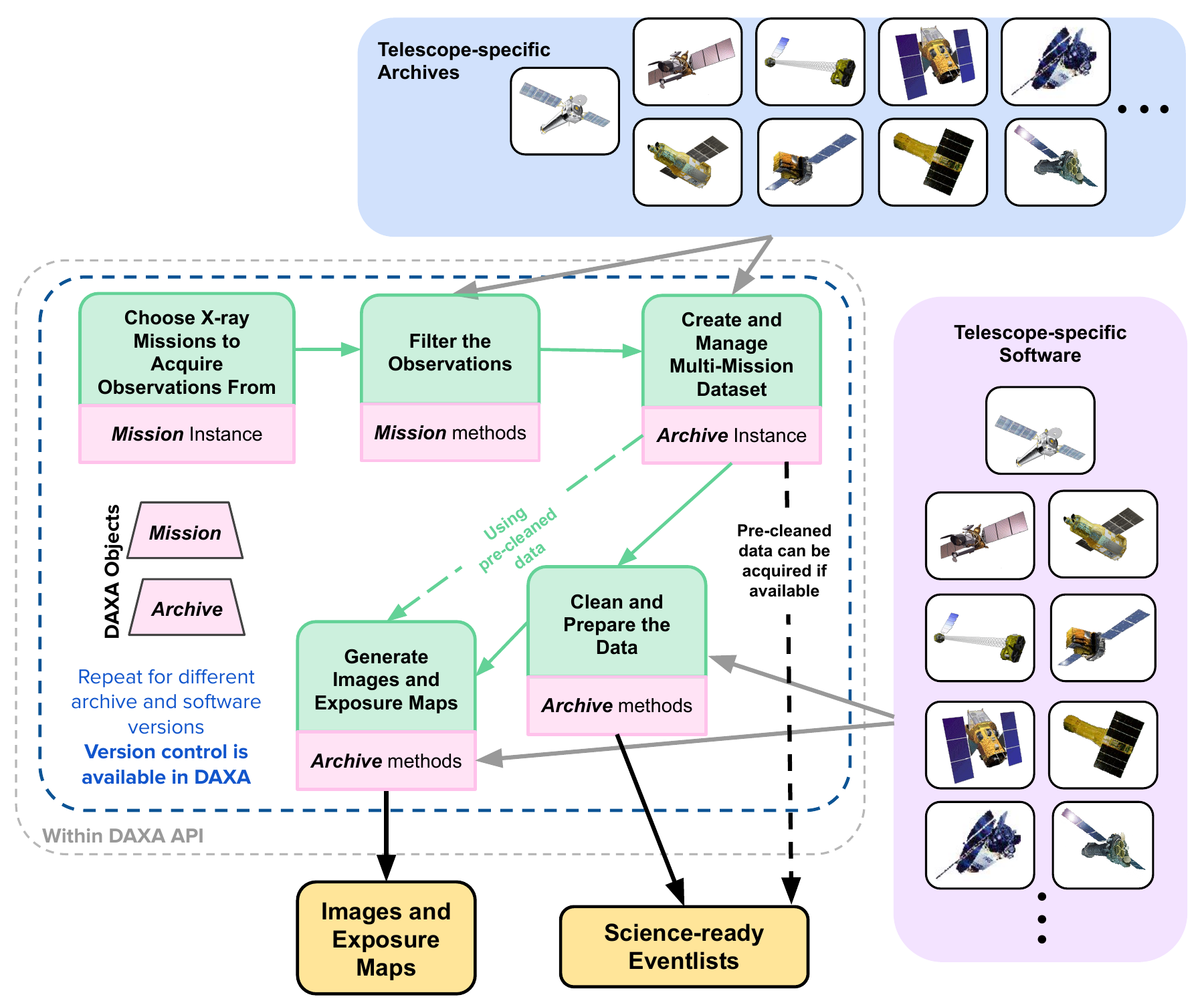}
\caption{A flowchart showing a brief overview of the \textsc{Daxa}
workflow. We indicate the different ways that \textsc{Daxa} can be used
to access, process, and use archival X-ray data. \label{fig:flowchart}}
\end{figure}

As such, X-ray data should be made as accessible as possible, both for
X-ray experts and non-specialists who may face barriers to entry; X-ray
data can be particularly intimidating to those astronomers who have not
used it before, though their research may benefit from a high-energy
view. Difficulty of use undermines the open-source nature of X-ray
astronomy data, which organisations such as the European Space Agency
(ESA) and the High Energy Astrophysics Science Archive Research Center
(HEASARC) have gone to great lengths to build. This may limit the reach
and scientific impact of X-ray telescopes; we should seek to maximise
the user of X-ray data, both to support X-ray astronomy through the
`X-ray desert', and to persuade funding bodies of the great need for
further X-ray telescopes.

We build on ESA and HEASARC's success and make the data more accessible
by providing a normalised interface to different backend software
packages and datasets, allowing for the easy processing of X-ray data to
a scientifically useful state; this is in addition to the ability to
download pre-processed data from many of the data archives. Through
\textsc{Daxa}, most X-ray observatory archives are accessible through a
single unified interface available in a programming language that is
ubiquitous in astronomy (Python); locally searching for data relevant to
a particular sample gives us the opportunity to better record and share
the exact search parameters, through a Jupyter notebook for instance.

\section{Features}\label{features}

\textsc{Daxa} contains two types of Python class: mission classes and
the archive class (see \autoref{fig:flowchart} for a schematic of the
structure of the module). Mission classes directly represent a telescope
or survey (for instance there are separate classes for pointed and
survey observations taken by \emph{ROSAT}
(\citeproc{ref-rosat}{Truemper, 1982}), as the characteristics of the
data are quite different), and exist to provide a Python interface with
the current telescope observation database. Such mission classes allow
the user to easily identify data relevant to their objects of interest
with various filtering methods (it is also possible to download the
entire archive of a telescope); these include filtering on spatial
position (determining whether a coordinate of interest is within the
field-of-view), filtering on the time of the observation (also filtering
on whether a specific coordinate was observed at a specific time, for
whole samples with different coordinates and times of interest), and
filtering on specific observation identifiers (ObsIDs) if they are
already known. Each mission class has some knowledge of the
characteristics of the telescope it represents (such as the
field-of-view) to make observation filtering easier. The user can also
select a subset of instruments, if the telescope has more than one, to
exclude any that may not contribute to their analysis.

Once a set of relevant observations have been identified, for either a
single mission or a set of missions, a \textsc{Daxa} data archive can be
declared. When a user declares a \textsc{Daxa} archive, the selected
data are automatically downloaded from the various telescope datasets,
and then ingested and organised so that they can be managed through the
\textsc{Daxa} interface. We have also implemented user-friendly,
multi-threaded, data preparation and cleaning routines for some
telescopes (\emph{XMM} and \emph{eROSITA} in particular, though more
will be added); fine control of the parameters that configure these
processes is retained, but default behaviours can be used if the user is
unfamiliar with the minutiae of X-ray data preparation. Another key
benefit of reducing data with \textsc{Daxa} is the easy access to data
logs through its interface, in case of suspected problems during the
reduction processes. The module is also capable of safely handling
processing failures, recording at which processing step the failure
occurred for a particular ObsID.

All of this information is retained permanently, not just while the
initial \textsc{Daxa} processes are running. Any \textsc{Daxa} archive
can be loaded back in after the initial processing, once again providing
access to the stored logs, and processing information. At this point the
archives can also be updated, either by searching for new data from the
existing missions, adding data from a different mission, or
re-processing specific observations to achieve more scientifically
useful data. Any such change will be recorded in the archive history, so
that the data archive can have a specific version that refers to its
exact state at any given time; this version can be referred to in
published work using the data archive. Each data archive is also capable
of creating a file that other \textsc{Daxa} users can import, and which
will recreate the data archive by downloading the same data, and
processing it in the same way; this renders making fully processed, and
large, X-ray data files available with a piece of research unnecessary.
This feature in particular can be used to further one of the tenets of
open-source science - reproducibility.

\section{Existing software packages}\label{existing-software-packages}

There are no direct analogues to our module, though we must acknowledge
the many pieces of software (and data archives), that greatly
facilitated the development of \textsc{Daxa}. Data access is made
possible primarily by the HEASARC data archive, though the Astroquery
(\citeproc{ref-astroquery}{Ginsburg et al., 2019}) module is also used.
HEASARC provides an online interface to query their data archive, which
has similar functionality to some of the filtering methods of mission
classes in \textsc{Daxa} (though we provide slightly more functionality
in that regard), and they provide Python SQL examples to access the
data, but none of the data management and cleaning functionality that we
include.

\textsc{Daxa} also builds on the various telescope-specific software
packages to perform data preparation and cleaning. Particularly
important are the \emph{XMM} Science Analysis System (SAS; Gabriel et
al. (\citeproc{ref-sas}{2004})) and the complementary extended SAS
(eSAS; Snowden \& Kuntz (\citeproc{ref-esascook}{2011})) packages, which
allow us to provide simple Python interfaces to the complex, multi-step,
processes that are required to prepare raw \emph{XMM} data for
scientific use. The analogous \emph{eROSITA} Science Analysis Software
System (eSASS; Predehl et al. (\citeproc{ref-erosita}{2021})) must also
be mentioned, as it provides the tools needed to reduce and prepare
\emph{eROSITA} data. In this vein we must also acknowledge the HEASoft
package, which is almost ubiquitous in X-ray data analyses, and is used
by both SAS and eSASS.

Another related software package is the other module in our open-source
X-ray astronomy ecosystem, X-ray: Generate and Analyse (\textsc{Xga};
Turner et al. (\citeproc{ref-xga}{2022})) - it exists to analyse large
samples of sources using large sets of X-ray data. \textsc{Daxa} is
designed to go hand-in-hand with \textsc{Xga}, as it will build and
manage the kind of dataset required for \textsc{Xga} to attain maximum
usefulness. We emphasise that such datasets do not \emph{have} to be
analysed with \textsc{Xga} however.

We have created a one-stop-shop for downloading and processing archival
X-ray data, making it more accessible and user-friendly, particularly
for non-specialists. \textsc{Daxa} is greater than the sum of its parts,
but is only possible because of the existing software packages it builds
upon; we hope that it only enhances the value that astrophysicists
derive from the other software we have mentioned.

\section{Future Work}\label{future-work}

The most significant new features implemented in \textsc{Daxa} will be
new mission classes added when new X-ray telescope archives become
available, or one of the existing missions that we have not yet
implemented is added (for instance \emph{XMM} observations taken whilst
slewing). We will also seek to include support for more
telescope-specific cleaning methods taken from their backend software;
additionally we wish to implement our own generic processing and
cleaning techniques where possible, applicable to multiple missions. We
also aim to include source detection capabilities; specifically
techniques that are generally applicable to multiple missions whilst
taking into account instrument-specific effects.

\section{Acknowledgements}\label{acknowledgements}

DT and MD are grateful for support from the National Aeronautic and
Space Administration Astrophysics Data Analysis Program
(NASA80NSSC22K0476). KR and PG acknowledge support from the UK Science and Technology Facilities Council via grants ST/T000473/1 and ST/X001040/1. JP
acknowledges support from the UK Science and Technology Facilities
Council via grants ST/X508822/1. DT is grateful to Amanda Witwer for comments on the structure and content of this paper.

\section*{References}\label{references}
\addcontentsline{toc}{section}{References}

\phantomsection\label{refs}


\begin{CSLReferences}{1}{0}
\bibitem[\citeproctext]{ref-cygx1andfriends}
Bowyer, S., Byram, E. T., Chubb, T. A., \& Friedman, H. (1965). {Cosmic
X-ray Sources}. \emph{Science}, \emph{147}(3656), 394--398.
\url{https://doi.org/10.1126/science.147.3656.394}

\bibitem[\citeproctext]{ref-sas}
Gabriel, C., Denby, M., Fyfe, D. J., Hoar, J., Ibarra, A., Ojero, E.,
Osborne, J., Saxton, R. D., Lammers, U., \& Vacanti, G. (2004). {The
XMM-Newton SAS - Distributed Development and Maintenance of a Large
Science Analysis System: A Critical Analysis}. In F. Ochsenbein, M. G.
Allen, \& D. Egret (Eds.), \emph{Astronomical data analysis software and
systems (ADASS) XIII} (Vol. 314, p. 759).

\bibitem[\citeproctext]{ref-lynx}
Gaskin, J. A., Swartz, D. A., Vikhlinin, A., Özel, F., Gelmis, K. E.,
Arenberg, J. W., Bandler, S. R., Bautz, M. W., Civitani, M. M.,
Dominguez, A., Eckart, M. E., Falcone, A. D., Figueroa-Feliciano, E.,
Freeman, M. D., Günther, H. M., Havey, K. A., Heilmann, R. K., Kilaru,
K., Kraft, R. P., \ldots{} Zuhone, J. A. (2019). {Lynx X-Ray
Observatory: an overview}. \emph{Journal of Astronomical Telescopes,
Instruments, and Systems}, \emph{5}, 021001.
\url{https://doi.org/10.1117/1.JATIS.5.2.021001}

\bibitem[\citeproctext]{ref-swift}
Gehrels, N., Chincarini, G., Giommi, P., Mason, K. O., Nousek, J. A.,
Wells, A. A., White, N. E., Barthelmy, S. D., Burrows, D. N., Cominsky,
L. R., Hurley, K. C., Marshall, F. E., Mészáros, P., Roming, P. W. A.,
Angelini, L., Barbier, L. M., Belloni, T., Campana, S., Caraveo, P. A.,
\ldots{} Zhang, W. W. (2004). {The Swift Gamma-Ray Burst Mission}.
\emph{611}(2), 1005--1020. \url{https://doi.org/10.1086/422091}

\bibitem[\citeproctext]{ref-coup}
Getman, K. V., Flaccomio, E., Broos, P. S., Grosso, N., Tsujimoto, M.,
Townsley, L., Garmire, G. P., Kastner, J., Li, J., Harnden, Jr., F. R.,
Wolk, S., Murray, S. S., Lada, C. J., Muench, A. A., McCaughrean, M. J.,
Meeus, G., Damiani, F., Micela, G., Sciortino, S., \ldots{} Feigelson,
E. D. (2005). {Chandra Orion Ultradeep Project: Observations and Source
Lists}. \emph{160}(2), 319--352. \url{https://doi.org/10.1086/432092}

\bibitem[\citeproctext]{ref-theOG}
Giacconi, R., Gursky, H., Paolini, F. R., \& Rossi, B. B. (1962).
{Evidence for x Rays From Sources Outside the Solar System}.
\emph{9}(11), 439--443. \url{https://doi.org/10.1103/PhysRevLett.9.439}

\bibitem[\citeproctext]{ref-astroquery}
Ginsburg, A., Sipőcz, B. M., Brasseur, C. E., Cowperthwaite, P. S.,
Craig, M. W., Deil, C., Guillochon, J., Guzman, G., Liedtke, S., Lian
Lim, P., Lockhart, K. E., Mommert, M., Morris, B. M., Norman, H.,
Parikh, M., Persson, M. V., Robitaille, T. P., Segovia, J.-C., Singer,
L. P., \ldots{} a subset of the astropy collaboration. (2019).
{astroquery: An Astronomical Web-querying Package in Python}.
\emph{157}, 98. \url{https://doi.org/10.3847/1538-3881/aafc33}

\bibitem[\citeproctext]{ref-xest}
Güdel, M., Briggs, K. R., Arzner, K., Audard, M., Bouvier, J.,
Feigelson, E. D., Franciosini, E., Glauser, A., Grosso, N., Micela, G.,
Monin, J.-L., Montmerle, T., Padgett, D. L., Palla, F., Pillitteri, I.,
Rebull, L., Scelsi, L., Silva, B., Skinner, S. L., \ldots{} Telleschi,
A. (2007). {The XMM-Newton extended survey of the Taurus molecular cloud
(XEST)}. \emph{468}(2), 353--377.
\url{https://doi.org/10.1051/0004-6361:20065724}

\bibitem[\citeproctext]{ref-nustar}
Harrison, F. A., Craig, W. W., Christensen, F. E., Hailey, C. J., Zhang,
W. W., Boggs, S. E., Stern, D., Cook, W. R., Forster, K., Giommi, P.,
Grefenstette, B. W., Kim, Y., Kitaguchi, T., Koglin, J. E., Madsen, K.
K., Mao, P. H., Miyasaka, H., Mori, K., Perri, M., \ldots{} Urry, C. M.
(2013). {The Nuclear Spectroscopic Telescope Array (NuSTAR) High-energy
X-Ray Mission}. \emph{770}(2), 103.
\url{https://doi.org/10.1088/0004-637X/770/2/103}

\bibitem[\citeproctext]{ref-xmm}
Jansen, F., Lumb, D., Altieri, B., Clavel, J., Ehle, M., Erd, C.,
Gabriel, C., Guainazzi, M., Gondoin, P., Much, R., Munoz, R., Santos,
M., Schartel, N., Texier, D., \& Vacanti, G. (2001). {XMM-Newton
observatory. I. The spacecraft and operations}. \emph{365}, L1--L6.
\url{https://doi.org/10.1051/0004-6361:20000036}

\bibitem[\citeproctext]{ref-clusterdisco}
Kellogg, E., Gursky, H., Leong, C., Schreier, E., Tananbaum, H., \&
Giacconi, R. (1971). {X-Ray Observations of the Virgo Cluster, NGC 5128,
and 3c 273 from the UHURU Satellite}. \emph{165}, L49.
\url{https://doi.org/10.1086/180714}

\bibitem[\citeproctext]{ref-cavities}
McNamara, B. R., Wise, M. W., Nulsen, P. E. J., David, L. P., Carilli,
C. L., Sarazin, C. L., O'Dea, C. P., Houck, J., Donahue, M., Baum, S.,
Voit, M., O'Connell, R. W., \& Koekemoer, A. (2001). {Discovery of Ghost
Cavities in the X-Ray Atmosphere of Abell 2597}. \emph{562}(2),
L149--L152. \url{https://doi.org/10.1086/338326}

\bibitem[\citeproctext]{ref-qpedisco}
Miniutti, G., Saxton, R. D., Giustini, M., Alexander, K. D., Fender, R.
P., Heywood, I., Monageng, I., Coriat, M., Tzioumis, A. K., Read, A. M.,
Knigge, C., Gandhi, P., Pretorius, M. L., \& Ag\'{i}s-Gonz\'{a}lez, B. (2019).
{Nine-hour X-ray quasi-periodic eruptions from a low-mass black hole
galactic nucleus}. \emph{573}(7774), 381--384.
\url{https://doi.org/10.1038/s41586-019-1556-x}

\bibitem[\citeproctext]{ref-athena}
Nandra, K., Barret, D., Barcons, X., Fabian, A., den Herder, J.-W.,
Piro, L., Watson, M., Adami, C., Aird, J., Afonso, J. M., Alexander, D.,
Argiroffi, C., Amati, L., Arnaud, M., Atteia, J.-L., Audard, M.,
Badenes, C., Ballet, J., Ballo, L., \ldots{} Zhuravleva, I. (2013). {The
Hot and Energetic Universe: A White Paper presenting the science theme
motivating the Athena+ mission}. \emph{arXiv e-Prints}, arXiv:1306.2307.
\url{https://arxiv.org/abs/1306.2307}

\bibitem[\citeproctext]{ref-xrayirrexo}
Poppenhaeger, K., Ketzer, L., \& Mallonn, M. (2021). {X-ray irradiation
and evaporation of the four young planets around V1298 Tau}.
\emph{500}(4), 4560--4572. \url{https://doi.org/10.1093/mnras/staa1462}

\bibitem[\citeproctext]{ref-erosita}
Predehl, P., Andritschke, R., Arefiev, V., Babyshkin, V., Batanov, O.,
Becker, W., Böhringer, H., Bogomolov, A., Boller, T., Borm, K.,
Bornemann, W., Bräuninger, H., Brüggen, M., Brunner, H., Brusa, M.,
Bulbul, E., Buntov, M., Burwitz, V., Burkert, W., \ldots{} Yaroshenko,
V. (2021). {The eROSITA X-ray telescope on SRG}. \emph{647}, A1.
\url{https://doi.org/10.1051/0004-6361/202039313}

\bibitem[\citeproctext]{ref-esascook}
Snowden, S. L., \& Kuntz, K. D. (2011). \emph{Cookbook for analysis
procedures for XMM-newton EPIC MOS observations of extended objects and
the diffuse background}.

\bibitem[\citeproctext]{ref-rosat}
Truemper, J. (1982). {The ROSAT mission}. \emph{Advances in Space
Research}, \emph{2}(4), 241--249.
\url{https://doi.org/10.1016/0273-1177(82)90070-9}

\bibitem[\citeproctext]{ref-xga}
Turner, D. J., Giles, P. A., Romer, A. K., \& Korbina, V. (2022). {XGA:
A module for the large-scale scientific exploitation of archival X-ray
astronomy data}. \emph{arXiv e-Prints}, arXiv:2202.01236.
\url{https://arxiv.org/abs/2202.01236}

\bibitem[\citeproctext]{ref-xrism}
XRISM Science Team. (2020). {Science with the X-ray Imaging and
Spectroscopy Mission (XRISM)}. \emph{arXiv e-Prints}, arXiv:2003.04962.
\url{https://arxiv.org/abs/2003.04962}

\bibitem[\citeproctext]{ref-miscen}
Zhang, Y., Jeltema, T., Hollowood, D. L., Everett, S., Rozo, E., Farahi,
A., Bermeo, A., Bhargava, S., Giles, P., Romer, A. K., Wilkinson, R.,
Rykoff, E. S., Mantz, A., Diehl, H. T., Evrard, A. E., Stern, C., Gruen,
D., von der Linden, A., Splettstoesser, M., \ldots{} DES Collaboration.
(2019). {Dark Energy Surveyed Year 1 results: calibration of cluster
mis-centring in the redMaPPer catalogues}. \emph{487}(2), 2578--2593.
\url{https://doi.org/10.1093/mnras/stz1361}

\end{CSLReferences}
\end{document}